\documentclass[aps,pre,twocolumn,groupedaddress,showpacs,floatfix]{revtex4}
\usepackage{graphicx}
\usepackage{dcolumn}
\usepackage{bm}
\usepackage{amssymb}
\usepackage{amsmath}
\usepackage{epsfig}

\begin{document}

\title{Generation of dynamic structures in nonequilibrium reactive bilayers}

\author{Ramon Reigada$^{1},$ Javier Buceta$^{2}$ and Katja Lindenberg$^{3} $}
\affiliation{$^{(1)}$Departament de Qu\'{\i}mica-F\'{\i}sica, Universitat de 
Barcelona,
Avda. Diagonal 647, 08028 Barcelona, Spain\\
$^{(2)}$Centre de Recerca en Qu\'{\i}mica Te\`orica (CeRQT),
Parc Cient\'{\i}fic de Barcelona, Campus Diagonal - Universitat de
Barcelona, Edifici Modular, C/Josep Samitier 1-5, 08028 Barcelona,
Spain\\
$^{(3)}$Department of Chemistry and Biochemistry 0340, and Institute for
Nonlinear Science, University of California, San Diego,
9500 Gilman Drive, La Jolla, CA 92093, USA}

\begin{abstract}
We present a nonequlibrium approach for the study of a flexible
bilayer whose two components induce
distinct curvatures. In turn, the two components are interconverted by an
externally promoted reaction. Phase separation of the two
species in the surface results in the growth of domains characterized by
different local composition and curvature
modulations. This domain growth is limited by the effective mixing due
to the interconversion reaction, leading to a finite characteristic domain
size.  In addition to these effects, first introduced in our earlier
work [Phys. Rev. E {\bf 71}, 051906 (2005)],
the important new feature is the assumption that the reactive
process actively affects the local curvature of the
bilayer.  Specifically, we suggest that a 
force energetically activated by external sources
causes a modification of the shape of the membrane at the reaction site.
Our results show the appearance of a rich and robust
dynamical phenomenology that includes 
the generation of traveling and/or oscillatory patterns.
Linear stability analysis, amplitude equations and numerical
simulations of the model kinetic equations confirm the occurrence of
these spatiotemporal behaviors in nonequilibrium reactive bilayers.
\end{abstract}

\pacs{87.16.Dg, 07.05.Tp, 82.45.Mp}
\maketitle

\section{Introduction}
\label{intro}

Due to hydrophobic repulsions, amphiphilic molecules such as lipids
spontaneously aggregate in water
to form bilayer membranes. These bilayers typically exhibit an
in-plane fluidlike nature, and are highly flexible surfaces so they
can display a large variety of conformations. 
The experimental study of equilibrium lipid bilayers has attracted
a great deal of attention in the past decades~\cite{lipnat,lip}.
Issues such as membrane conformational behavior, 
shape fluctuations, fusion and fission, and phase segregation in
multicomponent bilayers
have been extensively investigated~\cite{lipnat,lip,seif}. 
Among many other manipulation techniques, the fabrication of synthetic
giant vesicles and planar lipid membranes~\cite{blm} as well as 
micropipet aspiration~\cite{evans} have been used to perform these
studies, leading to a fairly detailed current understanding of
equilibrium membranes.

Parallel to the experimental advances, theoretical modeling 
of equilibrium flexible membranes has also 
come a long way. Typical theoretical approaches are based on
the search for minimum energy conformations
according to the Canham-Helfrich bending elasticity
description~\cite{canham,nuovo}
of a tensionless membrane, plus (optionally) the corresponding
area-difference contributions for closed surfaces~\cite{adif}.
More recently, it has been recognized that internal degrees of freedom
such as chemical composition
can crucially influence the shape of the bilayer, especially when
dealing with multicomponent membranes undergoing phase separation.
In particular, models containing a local coupling between composition
and curvature have been developed~\cite{and,sun,goz,dyn3,dyn2,dyn1},
resulting in an interesting outcome, namely, the appearance of phase
separating domains that grow and adopt distinct membrane curvatures.
This is understood as the initiation mechanism for budding phenomena
in real membranes.  Kinetic mesoscopic schemes, Monte Carlo approaches,
and a variety of discrete dynamic algorithms have been proposed on
this basis, and have mainly been devoted to the study of the
equilibrium behavior and the dynamics toward equilibrium of
flexible bilayers.

The ultimate motivation for the study of lipid bilayers lies in
their likeness to cell membranes. A biological membrane
is a complex mixture of lipids, sterols and proteins that represents
the main structural component of the cell
architecture. Rather than an inert static boundary, a membrane has to
be viewed as a dynamical surface directly and actively
involved in many biological cell processes~\cite{hous}. In addition to their
compositional complexity, cell membranes are also continuously subjected
to nonequilibrium driving forces, chemical gradients, and energy fluxes,
so it should be recognized that nonequilibrium behavior may underlie
many aspects of their dynamics~\cite{lenon,exert1,exert2}. Therefore,
although experimental and theoretical studies on thermally
equilibrated membranes have had considerable success and are a
good starting point, nonequilibrium approaches are
fundamental for a more accurate understanding of the dynamical
properties of both natural membranes and synthetic lipid bilayers.

Excellent work on nonequilibrium lipid bilayers has recently been 
presented by the experimental groups of J. Prost
and H.-G. D\"{o}bereiner. J. Prost et al. have studied the effects of
active proteins inserted in a vesicle membrane.
Acting as ion pumps externally activated by light, these proteins
provide a nonthermal energy source that directly
affects membrane shape behavior~\cite{prost4}. 
As an example of another nonequilibrium source, 
H.-G. D\"{o}bereiner et al. investigated the morphological transformations
of a vesicle membrane due to a photoinduced chemical reaction that
modifies the local lipid composition of the bilayer~\cite{petrov}.
Some progress has also been made on the modeling front.
For example, parallel to their experimental findings, a novel nonequilibrium
scheme has been proposed by J. Prost et al. that successfully
explains the observed experimental phenomenology~\cite{prost2,prost3}.
In addition to this proposal, other nonequilibrium situations have
been modeled very recently, such as those of membranes
confined between parallel plates~\cite{wada},
membranes with active inclusions~\cite{chen}, and bilayers near
repulsive walls~\cite{sum}. 

We are mainly interested in the nonequilibrium situation proposed
by H.-G. D\"{o}bereiner et al.,
where the bilayer's own lipid constituents are chemically
transformed, leading to changes in the membrane curvature.
The influence of lipid composition on the curvature of cell
membranes has recently been established in cell fusion processes
where high-curvature lipids play an important role~\cite{sci04}.
Chemically induced lipid modifications
are also known to be responsible for membrane shape transformations
involved in nervous synaptic processes such as the formation
of microvesicles that release the neurotransmitter to the gap
between two nerve cells~\cite{scales}.
Furthermore, in addition to biological membranes, an understanding of
the coupling between chemical reactions and the interfacial curvature is
essential in elucidating dynamical processes in artificial bilayers
that may be useful in nanotechnology applications.
As an example, self-assembly techniques based on biomineralization
can lead to hierarchically organized materials built from reactive
amphiphilic interfaces~\cite{biomin}.

Based on the nonequilibrium ingredients described above, 
in a previous paper~\cite{primer} we presented a model for a
two-component reactive membrane that exhibits a particular compositional
and morphological organization.
In that model the two lipid constituents have a distinctly different
shape, so that they are able to produce
distinct curvatures in the membrane. Moreover, they are
assumed to be thermodynamically immiscible, so they spontaneously
induce the development of phase separating domains. Additionally,
an externally induced reaction (promoted by a
nonequilibrium source such as a chemical flux, an applied light, etc.)
interconverts the two lipids.
The combination of these ingredients leads to stationary
patterns involving heterogeneous modulations of
composition and curvature. The novelty of the
resulting pattern organization is that although stationary,
the patterns are generated in a nonequilibrium context, that is,
they are actively maintained by the competition between
thermodynamic (equilibrium) phase separation and the environmentally
induced (nonequilibrium) reaction.
At thermal equilibrium, the low affinity between the two immiscible
components generates domains with a characteristic
composition and curvature. In the absence of reaction, these
domains grow indefinitely and coalesce
until complete segregation into two large domains is achieved.
However, the reactive process converts one lipid
component into the other, resulting in a large-scale mixing mechanism
that halts the growth of the segregated
structures when the mixing action balances the short-scale
ordering effect of phase separation. This balance leads to a stationary
state with patterns of a finite size.

In our previous model~\cite{primer}, we assumed that the nonthermal
energy contribution that promoted the lipid interconversion
reaction simply dissipates to the medium. In this paper we generalize
our original model to the case where such an energetic process
directly alters the membrane shape by exerting local forces
on it, that is, the induced reactive process locally 'kicks' the membrane.
Conformational changes of the membrane constituents due to
the reaction can reasonably be expected to
have mechanical effects on the membrane.
We model this effect in a very generic way, and will demonstrate that
the inclusion of this new ingredient is essential for the generation
of dynamic spatial organization. 
As a typical feature of soft-matter systems, robust spatiotemporal
phenomena such as those displayed in this paper are interesting in
their applicability to real bilayer membranes.

Although our ideas are inspired by phenomena observed in real
cellular systems, we hasten to disclaim a direct analogy between our
model and in vivo systems which are far more complex than our stripped
model.  We only wish to suggest that the mechanisms proposed herein
might play a role in real membranes in which the energy sources for the
local membrane-shape-altering processes might include ATP consumption,
the asborption of light, or any of a number of other possible
energy providing mechanisms.
We point out here and again later in the paper that the spatiotemporal
structures of interest require parameters that constrain the range of
applicability of the model. In particular,  we will see that elastic
properties (bending rigidity) play a central role, and
that the most likely candidates for the observation of our predictions
due to their flexibility
are in vitro lipid bilayers formed from single- and/or short-chain lipid
surfactants, especially those composed of surfactants of rather distinct
shapes that are interconverted by a reactive process that can be
controlled experimentally.  We discuss these and other possibilities
for experimental realization later in the paper.

The outline of this paper is as follows. Section~\ref{model} presents
our basic mesoscopic free energy description as well as
the derivation of the kinetic equations for the composition and
curvature variables in the proposed model.
The linear stability analysis and the amplitude equation
analysis are presented
in Sec.~\ref{analysis}, with the details of the derivation of the latter
relegated to the Appendix. Organized spatiotemporal behaviors
are predicted in some regions of parameter space provided that the parameter
describing the activity of the reactive process on the membrane dynamics
is sufficiently large.  In good agreement with the analytical
predictions, numerical simulations exhibiting traveling and/or
oscillating domains are shown in Sec.~\ref{numer}. Finally,
in Sec.~\ref{concl} we summarize the main conclusions of this paper.

\section{The Model}
\label{model}
In our earlier work we considered the simplest scenario where one
layer of the membrane (say the outer one) was composed of two
differently shaped lipids $A$ and $B$, whereas the other layer was
composed of a single component without any curvature effect.
We also disregarded
any lipid {\em flip-flop} exchange between the inner and outer layers.
Under these assumptions, we modeled the properties of the non-active
membrane as a two-dimensional surface characterized by the local
concentration difference $\phi$ between the $A$ and $B$ components and
the local extrinsic curvature $H$ of the surface.  At this point we do
not specifically identify the components $A$ and $B$ in any further
detail since this is to be seen as a schematic and highly streamlined
representation of any number of possible realizations.  Indeed, $A$ and
$B$ need not even be single compounds; for example, $A$ may be a group
of lipids or biomolecules some of which induce a particular curvature,
while $B$ may be another group that induces a different curvature.
$A$ and $B$ could be the isomers of an amphiphillic azobenzene
derivative, or groups of lipids as in Fig. 2 in~\cite{scales} or those
suggested in~\cite{sci04}, or simply a given lipid with its polar head
more or less charged as in the experiments in~\cite{petrov}.

The free energy functional in terms of the two order parameters was
assumed to be  
\begin{eqnarray}
{\mathcal{F}}= &&\int dx dy \left[ -\frac{\alpha}{2}
\phi^{2}+ \frac{\beta}{4} \phi^{4}+ \frac{\gamma}{2}
|{\mathbf{\nabla}} \phi| ^{2} \right. \nonumber\\
&& \left. + \frac{\kappa}{2}
( \nabla^2 h - \phi H_0)^2  \right].
\label{fgen}
\end{eqnarray}
The first three terms correspond to the typical Ginzburg-Landau
expansion that leads to phase separation when $\alpha$, $\beta$,
and $\gamma$ are all positive, with an equilibrium concentration
difference $\phi_{eq}=\pm \sqrt{\alpha / \beta}$ and a typical
interface length $\zeta = \sqrt{\gamma/\alpha}$. When $\alpha$ is
negative the components are completely miscible, leading to a
homogeneous state.  The last term is the
elastic energy contribution due to the rigidity of the membrane, and
$\kappa$ is the bending rigidity modulus. For simplicity we
have assumed that the spontaneous (equilibrium) curvature
$H_{sp}(\phi)$,
which reflects the shape asymmetry between the two lipid components, is
linear in the order parameter, $H_{sp}=\phi H_0$.  
In the Monge parametrization~\cite{geom}
a deformable surface is described by $(x,y,h(x,y))$, where $h(x,y)$ is
the displacement (height)
field for the local separation from the flat conformation. This
representation is valid for surfaces that are nearly flat with
only gradual variations of $h$, and allows the approximation
$H \approx \nabla^2 h$ that we have incorporated in the free energy
functional.  For self-assembled free membranes, the surface tension
contribution
($\frac{\sigma}{2}|{\mathbf{\nabla}} h| ^{2}$) in the free energy
can be neglected, and we have not included it in Eq.~(\ref{fgen}).

The kinetics of $\phi$ and $h$ are assumed to be driven by the free
energy functional, and by the chemical reaction that interconverts the
two lipid components, $A\rightleftarrows B$.  The effect of the reaction
is two-fold: it affects the concentration difference order
parameter $\phi$, and it also affects the shape of the membrane.
This last effect is our new contribution in this paper. 
The kinetics is thus a generalized version of that introduced
in~\cite{primer}: 
\begin{subequations}
\begin{eqnarray}
\frac{\partial \phi}{\partial t} &=& 
D \nabla^{2} \left[ \frac{\partial {\mathcal{F}}}{\partial \phi}
\right] - \Gamma (\phi - \phi_0),
\label{phivar}
\\
\frac{\partial h }{\partial t}&=&- \Lambda \frac{\delta {\mathcal{F}}}{\delta
h } + \Gamma (\phi-\phi_0)\xi.
\label{hvar}
\end{eqnarray}
\end{subequations}
The concentration difference $\phi$ follows a conserved
scheme~\cite{sancho} with diffusion coefficient $D$, augmented by
the reaction contributions.  The rate parameter
$\Gamma=k_++k_-$ and the stationary concentration difference parameter
$\phi_0=(k_--k_+)/(k_++k_-)$ are determined by the
forward and backward reaction rate constants $k_+$ and $k_-$. 
Note that we have assumed local overdamped relaxational dynamics for the
membrane height, that is, Rouse-like dynamics, where $\Lambda$ denotes
a mobility parameter proportional to the inverse of the typical
relaxatio time $\tau_h$.  This approximation is only valid when
the membrane is immersed in a high viscosity medium
and/or when the membrane is highly permeable~\cite{frey}.
When hydrodynamic effects can not be
neglected, a more realistic, albeit complex, description of the
membrane dynamics (Zimm-like dynamics) is required, where inertial
effects and a space dependent relaxation parameter are in
order~\cite{moldovan}.

\begin{figure}[htb]
\begin{center}
\includegraphics[width=7cm,angle=0]{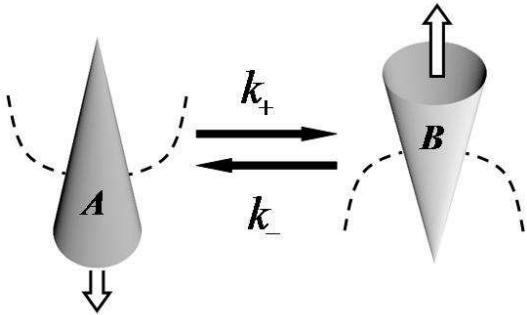}
\caption{Schematic representation of the model. Placed in one of
the layers of the membrane, the two constituents
constituents $A$ and $B$ induce
positive and negative curvatures, respectively (dashed lines). The
species are interconverted
by means of a nonequilibrium reaction: $k_+$ and $k_-$ are the forward
and backward rate
constants, respectively. The mechanical effect of
the reactive process on the membrane shape is illustrated by the hollow
arrows.}
\label{figmodel2}
\end{center}
\end{figure}

The last term in Eq.~(\ref{hvar}) 
is based on the following hypothesis.
The reaction that interconverts
$A$ and $B$ can be understood as an isomerization-like
chemical transformation, involving a strong modification in the
shape of the membrane constituents. This process
implies the displacement of parts of these molecules that could
have a mechanical effect on the local membrane shape.
If additionally the process is strongly energetically activated by
an external energy source, it might exert
a local force on the membrane.
A simplifying approximation is to
assume that the forward and backward reactions
exert opposite forces on the membrane.  These forces are assumed to
act locally for a negligible period of time (the time needed to complete
the reaction is much shorter than any other
time scale of the system), in the same direction as the
preferred curvature of the reaction product component, that is, positive (outwards)
for $A \rightarrow B$, and negative (inwards) for
$B \rightarrow A$, see Fig.~\ref{figmodel2}.
This is modeled in a generic way \cite{gov} by the last term in the membrane height
kinetic equation, where $\xi$ is a parameter accounting for the strength of the
effect of the reaction on the shape of the membrane. Here we take the parameter
$\xi$ to have the same sign as $H_0$ (in Fig.~\ref{figmodel2} both are positive).
Active proteins are known to act as force centers when inserted in lipid
bilayers~\cite{prost4}, and other experimental studies~\cite{exert1,exert2} also
provide evidence that reactive processes may locally modify the membrane
shape in red blood cells. However, we must recognize ours as an experimentally verifiable
conjecture of what might happen at the lipid component level since
the cited experimental evidence only corresponds
to chemical processes involving much larger molecules
(proteins). 

We write Eqs.~(\ref{phivar}) and (\ref{hvar}) in detail in
terms of dimensionless
parameters where the energy is measured in units of the thermal energy
$k_BT$, the time in units of $\tau_h$, and length in units of
$\sqrt{D\tau_h}$: 
\begin{subequations}
\begin{eqnarray}
\label{kineqa}
\frac{\partial \phi}{\partial t} &=& 
(\kappa H_0^{2} - \alpha) \nabla^{2} \phi + 3 \beta \phi^{2}\nabla^{2}\phi
+ 6 \beta \phi
|{\mathbf{\nabla}} \phi|^{2} \nonumber\\
&&~~-\gamma \nabla^{4}\phi -\kappa H_0 \nabla^{4}h
 - \Gamma \left( \phi - \phi_0 \right), \\
\frac{\partial h }{\partial t}&=&- \kappa \nabla^{4}h
+\kappa H_0 \nabla^{2} \phi + \Gamma (\phi-\phi_0)\xi.
\label{kineqb}
\end{eqnarray}
\end{subequations}

In the next two sections we show analytically and confirm numerically
that the new force term can have profound effects on the instability
and pattern formation properties of the model.  In particular, this term
may lead to dynamical patterns, whereas in its absence only stationary
patterns are possible.

\section{Analytical treatments}
\label{analysis}

The stationary uniform state corresponds to
$\overline{\phi}=\phi_0$ and arbitrary $\overline{h}$.
The linear stability of these uniform solutions is tested by adding
small plane-wave perturbations of wave numer $q$ to the uniform state and
linearizing Eqs.~(\ref{kineqa}) and (\ref{kineqb}) in these
perturbations.  This procedure leads to the $2\times 2$ linearization
matrix ${\mathcal{L}}$ with the following coefficients,
\begin{eqnarray}
\nonumber {\mathcal{L}}_{11} &=& -q^2
\left[ \left( \kappa H_0^2 -\alpha +3 \beta \phi_0^2 \right)
+\gamma q^2  \right] - \Gamma \\
\nonumber {\mathcal{L}}_{12} &=& -\kappa H_0 q^4\\
\nonumber {\mathcal{L}}_{21} &=& -\kappa H_0 q^2 + \Gamma \xi \\
{\mathcal{L}}_{22} &=& -\kappa q^4.
\label{coefl}
\end{eqnarray}
The eigenvalues $\omega_q$ of the Jacobian associated with the
matrix ${\mathcal{L}}$ correspond to the linear growth rates
of the perturbations.

The solutions of the eigenvalue problem are given by
$\omega_q=\frac{1}{2}\left(Tr[{\mathcal{L}}]
\pm \sqrt{\Delta[{\mathcal{L}}]}\right)$,
where $\Delta[{\mathcal{L}}]=Tr[{\mathcal{L}}]^2
-4Det[{\mathcal{L}}]$.
At the instability boundary,
${\rm Re}(\omega_q)$ vanishes for one finite wave number that is defined
as the first unstable mode.
If the imaginary part of $\omega_q$ is not zero at this wave number,
we have a wave bifurcation. The condition
for this bifurcation is obtained by requiring $Tr[{\mathcal{L}}]=0$
and $\Delta[{\mathcal{L}}]<0$. If the imaginary part of the growth rate
is zero, one has a Turing-like bifurcation. The condition for this
bifurcation is $Det[{\mathcal{L}}]=0$.

\begin{figure}[htb]
\begin{center}
\includegraphics[width=8cm ]{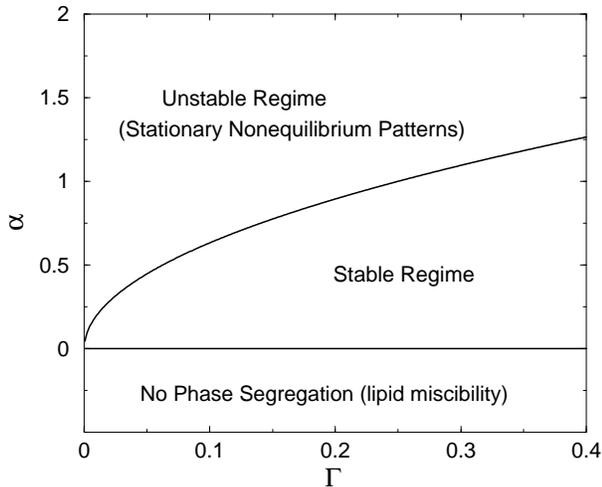}
\caption{Phase diagram for the case of immiscible components 
in the plane $(\alpha,\Gamma)$ for $\gamma=1$ and $\phi_0=0$
when there is no force on the membrane caused by the interconversion
reaction.  From~\cite{primer}.}
\label{figdiagfas}
\end{center}
\end{figure}

\begin{figure}[htb]
\begin{center}
\includegraphics[width=8cm ]{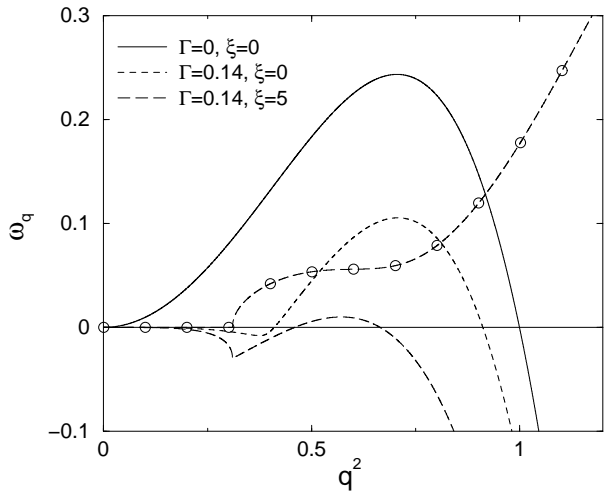}
\caption{Dispersion relation functions $\omega_q$ for different values of
$\Gamma$ and $\xi$.
The other parameters are held fixed at $\alpha=\gamma=1$, $\phi_0=0$,
$\kappa=0.5$ and $H_0=0.2$.  Only the case with
$\xi=5$
has a nonzero imaginary part of the dispersion relation (plotted with
symbols).}
\label{figomega}
\end{center}
\end{figure}

In the absence of the interconversion reaction it is of course well
known that immiscibility leads to complete phase separation. This occurs
because a continuous range of modes starting from $q=0$ is unstable, and
phase separation does not stop until there is complete segregation into
two large domains.  In our previous work~\cite{primer} we showed that the
interconversion reaction provides a mixing mechanism that stabilizes the
longest wavelength modes.  If the reaction rate parameter $\Gamma$ is
sufficiently large, 
\begin{equation}
\Gamma > \Gamma_1 =
\frac{\left( \alpha - 3 \beta \phi_{0}^2 \right)^2}{4\gamma},
\label{tbif}
\end{equation}
then mixing is complete, the instability disappears, and the bilayer
becomes stable and essentially flat.  If $\Gamma < \Gamma_1$, the
reaction no longer stabilizes all modes, but it does
stabilize the longest wave vector modes. The first mode to become
unstable is
\begin{equation}
q^2_1 = \frac{\alpha - 3 \beta \phi_{0}^2}{2 \gamma}
=\sqrt{\frac{\Gamma_1}{\gamma}}.
\label{q0t}
\end{equation}
The longest wavelength unstable mode now lies at the finite value
\begin{equation}
q_{min}^2=\frac{\alpha - 3 \beta \phi_{0}^2}{2\gamma}
- \frac{1}{2\gamma}\sqrt{(\alpha - 3 \beta \phi_{0}^2
)^2 -4\gamma \Gamma}.
\label{qpm}
\end{equation}
This means that the phase separation process stops when the separated
regimes are of characteristic size $\sim q_{min}^{-1}$.  The patterns are
Turing-like (stationary) because the imaginary part of the growth rate
is zero at the bifurcation point.  A typical phase
diagram found earlier for this case is shown in Fig.~\ref{figdiagfas}.
Note that the pattern size is independent of curvature parameters.
Curvature reduces the unstable mode growth rates (see
the $\xi=0$ curves in Fig.~\ref{figomega}), but without changing
either the characteristic
size of the patterns nor the marginal condition~(\ref{tbif}). The effect
of curvature is thus restricted to the kinetics of the
phase separation process, as has been extensively analyzed in~\cite{primer}. 

\begin{figure*}[htb]
\begin{center}
\includegraphics[width=8cm ]{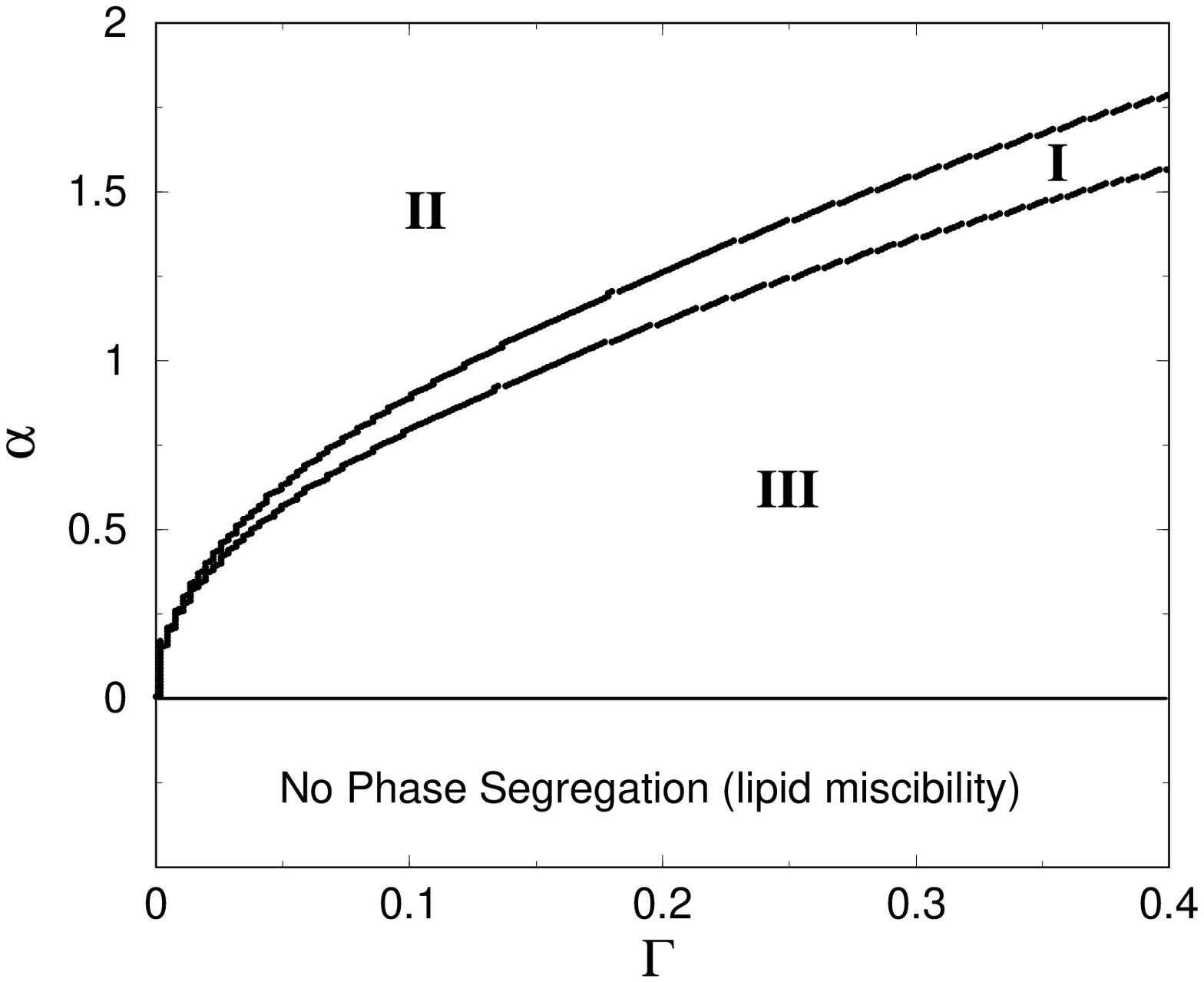}
\includegraphics[width=8cm ]{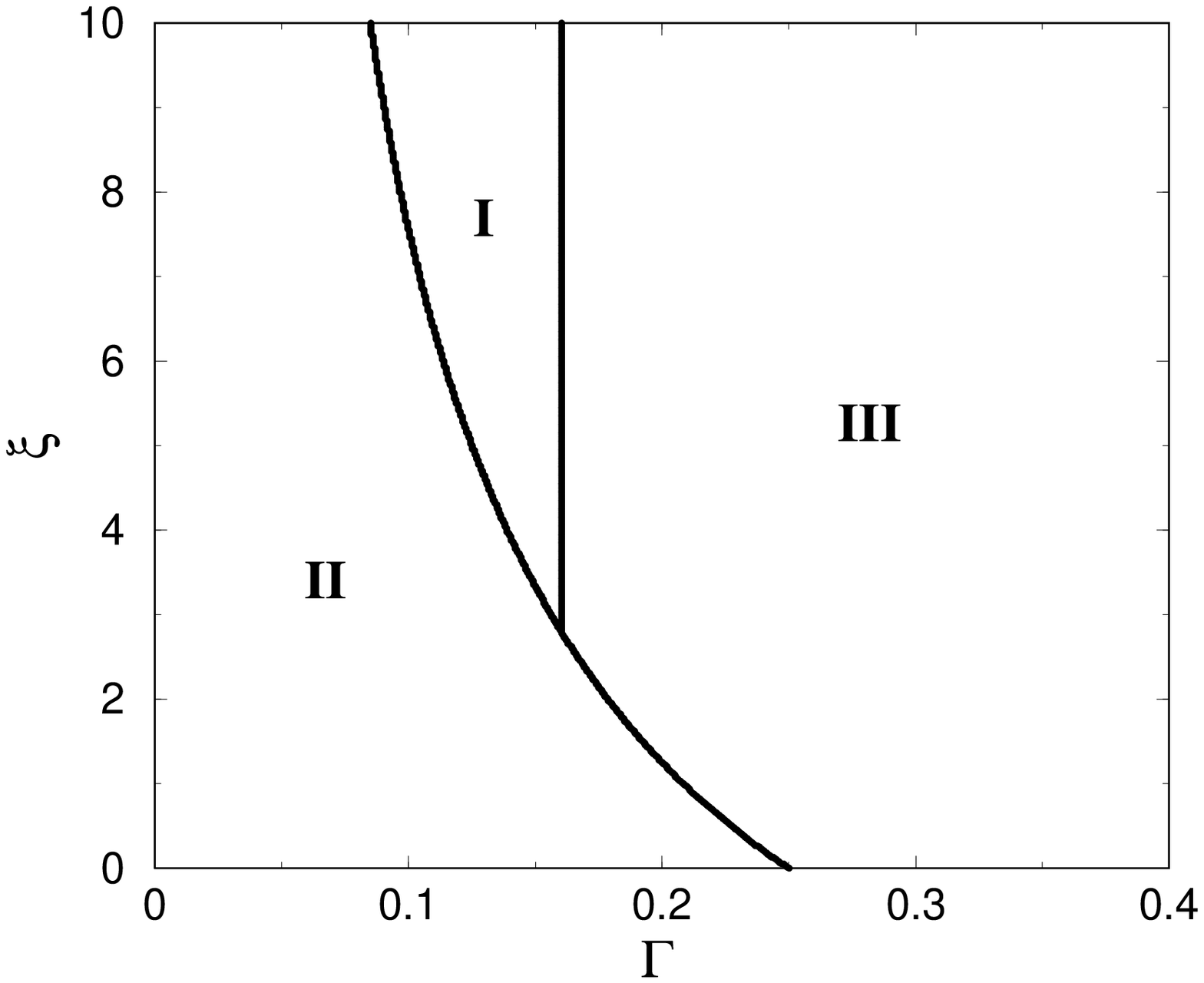}
\caption{Phase diagrams in the planes $(\alpha,\Gamma)$ for $\xi=5$,
and $(\xi,\Gamma)$ for $\alpha=1$.
The other parameter values are
$\gamma=1$, $\kappa=0.5$, $H_0=0.2$ and $\phi_0=0$.
The boundaries in these diagrams have been calculated numerically (see
Sec.~\ref{numer}).}
\label{figdiagfas2}
\end{center}
\end{figure*}

The new questions of interest here are the consequences of the
reaction--shape coupling in the kinetic equations. To place the analytic
results obtained in this section in context, we anticipate the numerical
results of the next section and exhibit two different views of a typical
phase diagram in Fig.~\ref{figdiagfas2}.
Regime III is the stable regime where the mixing due to the
interconversion is sufficiently strong to produce a homogeneous phase.
Here ``sufficiently strong" depends on the value of the parameter
$\xi$. Beyond this stable regime, there are now two instability
regimes.  One, called region II in the diagrams, is the Turing instability regime
leading to stationary nonequilibrium patterns, that is, the same sorts of patterns
observed in the absence of the reaction--shape coupling (upper region of
Fig.~\ref{figdiagfas})~\cite{primer}. 
The condition for the Turing-like bifurcation,
$Det[{\mathcal{L}}]=0$, obtained from the linear stability conditions
now is $\Gamma < \Gamma_2$, with
\begin{equation}
\Gamma_2 = \frac{\left( \alpha - 3 \beta \phi_{0}^2 \right)^2}
{4\gamma \left( 1+H_0 \xi\right)}.
\label{wbif}
\end{equation}
This is the curve bounding region II in the figure.
The first Turing-like unstable mode occurs at
\begin{equation}
q^2_2 = \frac{\alpha - 3 \beta \phi_{0}^2}{2 \gamma}
=\sqrt{\frac{\Gamma_2 \left( 1+H_0 \xi\right)}{\gamma}}.
\label{q0t2}
\end{equation}
The new regime, region I, is one in which one observes
dynamical (i.e., time-dependent)
patterns.  Provided the reaction parameter is sufficiently
small and the reaction--shape coupling parameter is sufficiently large,
the linear stability analysis in this
regime leads to wave bifurcation solutions, $Tr[{\mathcal{L}}]=0$
and $\Delta[{\mathcal{L}}]<0$.  The explicit conditions for these time
dependent solutions obtained from the linear stability conditions
are $\Gamma<\Gamma_0$, with
\begin{equation}
\Gamma_0 = \frac{\left( \alpha - 3 \beta \phi_{0}^2 -
\kappa H_0^2 \right)^2}{4\left( \gamma + \kappa \right)},
\label{condw}
\end{equation}
and $\xi>\xi_0$, with
\begin{equation}
\xi_0 =\frac{\kappa q^2_0}{\Gamma H_0}\left( H_0^2 + q^2_0
\right).
\label{xicond}
\end{equation}
The first unstable mode in the above expression now is
\begin{equation}
q^2_{0} = \frac{\alpha - 3 \beta \phi_{0}^2 -
\kappa H_0^2}{2 \left( \gamma+\kappa \right)}
=\sqrt{\frac{\Gamma_0}{\gamma + \kappa}}.
\label{q0w}
\end{equation}
The curves bounding region I in Fig.~\ref{figdiagfas2} arise from these
expressions.  A typical dispersion relation function $\omega(q)$
in this regime is shown in Fig.~\ref{figomega}. The function now has
a nonzero imaginary part as well as a real part, as appropriate for 
spatiotemporal pattern formation.
Typical spatiotemporal field configurations in the unstable regions of
the phase diagram can only be found numerically. We do so in the next
section.

Before doing so, however, we can go a step further in the analytic
approach to the problem by introducing a weakly nonlinear analysis based
on the amplitude equation formalism.
Such a nonlinear expansion is valid near the bifurcation threshold
to pattern formation and sheds light on the expected pattern configurations
and their dynamical properties. For simplicity, we restrict the
calculation of the amplitude equations to one spatial dimension. 
Nonetheless, we can readily deduce the possible arrangements in two
dimensions by an appropriate interpretion of the coupling terms.

The details of the derivation of the amplitude equations are presented
in the Appendix. The analysis reveals that a solution for equations
(\ref{kineqa}) and (\ref{kineqb}) near the bifurcation to pattern
formation reads
\begin{subequations}
\begin{eqnarray}
\phi(x,t)  &=&\phi_0  +A_+(x,t)\exp \left[i(q_0 x+\omega_0
t)\right]\nonumber\\
\label{ampa}
&&+A_-(x,t) \exp \left[i(q_0 x-\omega_0 t)\right] +c.c., \\
h\left( x,t\right) & =&B_+(x,t) \exp \left[ i(q_0 x+\omega_0
t)\right]\nonumber\\
&&+B_-(x,t)\exp \left[i(q_0 x-\omega_0 t)\right] +c.c.,
\label{ampb}
\end{eqnarray}
\end{subequations}
where $c.c.$ stands for the complex conjugate,
$\omega_0=\left\vert {\rm Im}(\omega _{q_0})\right\vert$,
and the amplitudes $A_{\pm }(x,t) $ and $B_{\pm }(x,t)$ satisfy
the equations
\begin{subequations}
\begin{eqnarray}
\label{apm}
\frac{\partial A_{\pm }}{\partial t}& =&(\Gamma_0-\Gamma)
A_{\pm }-\frac{3\beta q_0^2(\Gamma_0-\Gamma)}{\Gamma_0}\nonumber\\
& \times& \left( 1- \frac{24\beta \phi_0^2 q_0^2(8\kappa q_0^4
+i\omega_0)}{c_1}\right) A_{\pm }\left\vert A_{\pm }\right\vert
^2 \nonumber \\
&& -c_2 A_{\pm }\left\vert A_{\mp }\right\vert^2 \nonumber\\
&& +(3\beta \phi_0^2 +H_0^2\kappa +6\gamma q_0^2-\alpha)
\frac{\partial^2 A_{\pm }}{\partial x^2}  \nonumber\\
&&+6H_0\kappa q_0^2\frac{\partial^2 B_{\pm}}{\partial x^2}, \\
\frac{\partial B_{\pm}}{\partial t}& =&-\xi (\Gamma_0-\Gamma) A_{\pm}
\nonumber\\
&&+H_0\kappa \frac{\partial ^2 A_{\pm }}{\partial x^2}+
6\kappa q_0^2\frac{\partial ^2 B_{\pm}}{\partial x^2},
\label{bpm}
\end{eqnarray}
\end{subequations}
where
\begin{subequations}
\begin{eqnarray}
c_{1} &=&8\kappa q_0^4\left[\Gamma_0(1+H_0\xi) +4q_0^2(3\beta
\phi_0^2+4\gamma q_0^2-\alpha ) \right]  -2\omega_0^2 \nonumber\\ 
&&-i\omega_0\left[\Gamma_0+4q_0^2\left(3\beta
\phi_0^2 +H_0^2\kappa -\alpha +4q_0^2(\gamma +\kappa)
\right) \right), \nonumber \\
\\
c_{2} &=&\frac{6\beta q_0^2(\Gamma_0-\Gamma)}{\Gamma_0} \nonumber\\
&\times& \left( 1-\frac{24\beta \phi_0^2q_0^2}{\Gamma_0(1+H_0\xi)
+4q_0^2(3\beta \phi_0^2+4\gamma q_0^2-\alpha)}\right). \nonumber \\
\end{eqnarray}
\end{subequations}
Note that no pattern develops if $\Gamma >\Gamma _0$ since the
amplitudes then decay with time, $A_{\pm },B_{\pm }\rightarrow 0$.
Note also the coupling term between the
rightward, $A_-$, and leftward, $A_+$, traveling waves. This interaction is
responsible for generating either standing or traveling waves in the
membrane dynamics. Standing waves appear as a result of a
\emph{kink}-\emph{antikink} dynamics where both waves have the
same amplitude, $A_- = A_+$. Traveling waves require these
amplitudes to be different. Therefore, if the coefficient $c_2$
of this interaction term is small (strictly speaking, zero) we expect
the two amplitudes to be equal, so that
\begin{subequations}
\begin{eqnarray}
\phi (x,t) & =&\phi_0+4A_+(x,t)\cos(q_0x)\cos(\omega _0 t), \\
h(x,t) & =&4B_+(x,t)\cos(q_0x) \cos(\omega_0 t),
\end{eqnarray}
\end{subequations}
that is, the solution is a standing wave. On the other hand,
if $c_{2}$ is not zero, a traveling wave appears. Observe that if
$\phi_0=0$, that is, if the composition of the membrane is 50\%
lipid $A$ and 50\%
lipid $B$, then the $A_+\leftrightarrow A_-~$interaction term
is always relevant and no standing waves are possible. This scenario is
confirmed
by the numerical simulations described in the next section (cf.
Figs.~\ref{figtrav} and \ref{figperf}, where we see a composition-curvature
traveling wave in the membrane. For the parameters in that figure,
$c_{2}\approx 0.25$).
On the other hand, this restriction does not apply if the membrane
composition is such that $\phi_0\neq 0$ (cf. Figs.~\ref{figoscil} and
\ref{figmov}, where $\phi
_{0}=0.14$ and $c_{2}\approx $ $10^{-2}$; the simulations show that in
those cases there is a standing oscillatory contribution to the pattern).
As for the different
spatial configurations, one obtains either rolls or hexagons
depending on the symmetries of the system.
Thus, note that in Eqs.~(\ref{kineqa}) and (\ref{kineqb})
the inversion symmetry
\begin{equation}
\phi \rightarrow -\phi, \qquad h \rightarrow -h
\end{equation}
is satisfied only if $\phi_0=0$ and in that case roll-like
arrangements
are obtained. On the other hand, if $\phi _{0}\neq 0$ there is no
inversion symmetry, and the resulting structures are expected to be
hexagonal.

\section{Numerical Results}
\label{numer}

Representative numerical results are presented in this section,
showing good agreement with the predictions of the linear stability
and amplitude equation analyses.
We numerically solve Eqs.~(\ref{kineqa}) and (\ref{kineqb})
in a two-dimensional square lattice of $100 \times 100$ sites with
a mesh size $\Delta x=1$ and periodic boundary conditions.
The spatial derivatives are calculated by means of a simple centered
scheme, and a first-order Euler algorithm with
time step $\Delta t=10^{-4}$ is used for the temporal integration.
These coordinate and time steps insure good numerical accuracy.
Simulations are started from a slightly randomly perturbed
homogeneous distribution $\phi(\vec{r})=\overline{\phi}=\phi_0$ and
$h(\vec{r})=\overline{h}=0$.
In this section we present numerical results that correspond to
highly immiscible components (deep quench,
$\alpha=1$ and $\beta=1$, leading to an equilibrium value of
$\phi_{eq}=\pm 1$), and an interface thickness of the
order of the space discretization ($\zeta=1$, which leads to $\gamma=1$).
The bending rigidity modulus is taken equal to $0.5$ (in units of $k_BT$),
and very different shapes are implemented for the two constituents
by setting $H_0=0.2$. 

When dealing with mesosopic or coarse-grained modeling schemes one has to be
concerned about the applicability of the results. The limit on the
predictive power of these models is mainly determined by the
identification of the correct
time, length and energy scales accessible to the experiments. 
In our case, we have adimensionalized the kinetic equations in order to have
$D=\Lambda=k_BT=1$, so we can return dimensions to these equations by
choosing typical values for these parameters.
The diffusion coeficient is known to be in the range
$10^{-7}-10^{-8}$cm$^2$/s (for lipids
in a liquid-disordered phase membrane~\cite{hous}).
In the absence of the nonlinear and mechanical effects considered
in this paper, solvent hydrodynamic effects (Zimm dynamics)~\cite{seif}
can be incorporated with a renormalized height mobility
$\Lambda=\left( 4 \mu q \right)^{-1}$ in
Eq.~(\ref{hvar}) in Fourier space~\cite{frey}.
Here $\mu$ stands for the solvent kinematic viscosity, which is $1$cp 
for water at $20^{\circ}$C. For $\kappa=0.5 k_BT$ the linear
stability analysis shows typical unstable modes
at $q \approx q_0 \approx 0.5$, which corresponds to a pattern size
in the range of $2-20 \mu$m. Such a size may be
accessible in experiments on giant vesicles and planar bilayers.
However, when mechanical effects are included, a more elaborate
analysis such as that provided by dynamical renormalization
methods~\cite{frey} is in order.  A simple and plausible hypothesis
might be to include such effects by simply renormalizing {\em both}
$\Lambda$ and $\xi$ in Eq.~(\ref{hvar}), both with a $1/q$ decay in
Fourier space since one might expect a non-local viscous kernel to decay
as a power law with distance~\cite{moldovan}. 
In any case, a more detailed
consideration of this issue is beyond the scope of this work.

\begin{figure*}[htb]
\begin{center}
\leavevmode
\epsfxsize = 1.7in
\epsffile{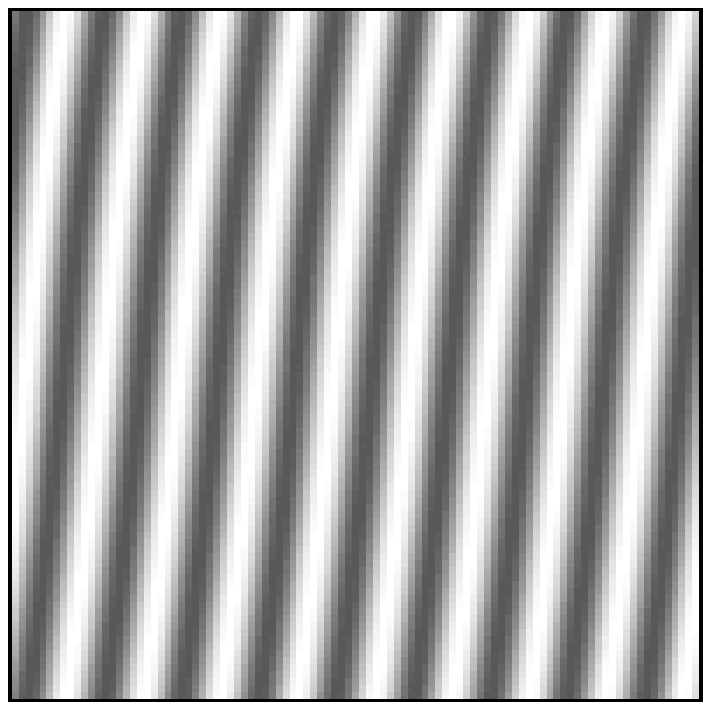}
\epsfxsize = 2.0in
\epsffile{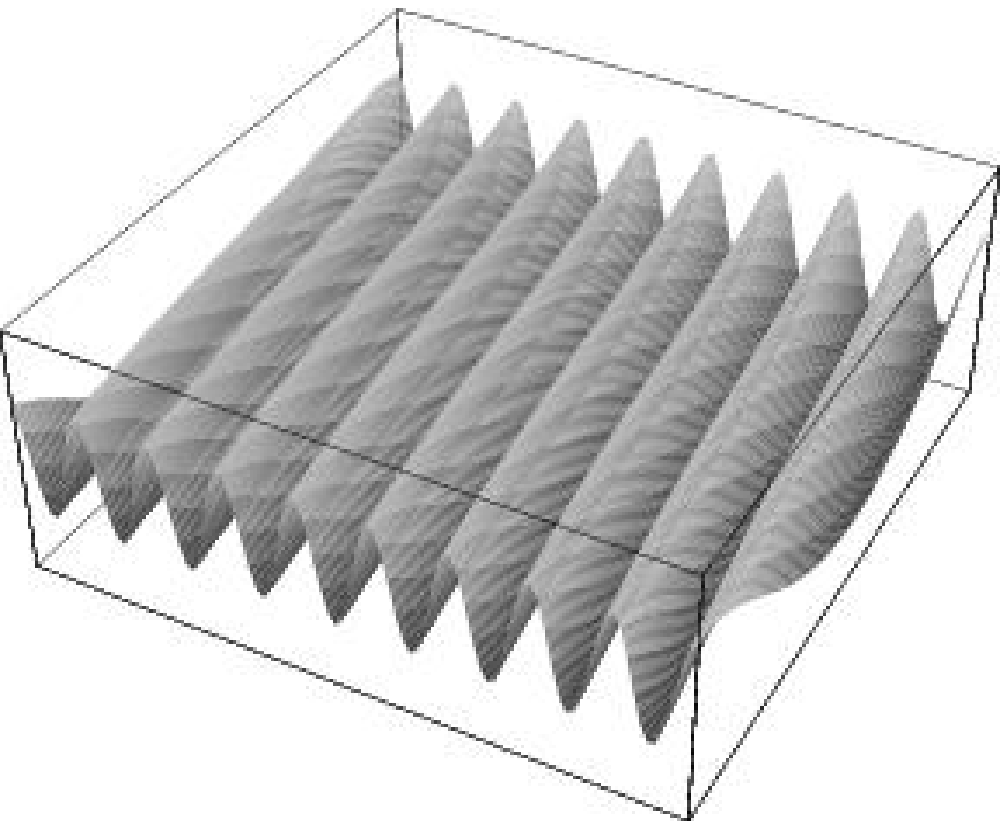} \\
\vspace{0.25in}
\epsfxsize = 2.2in
\epsffile{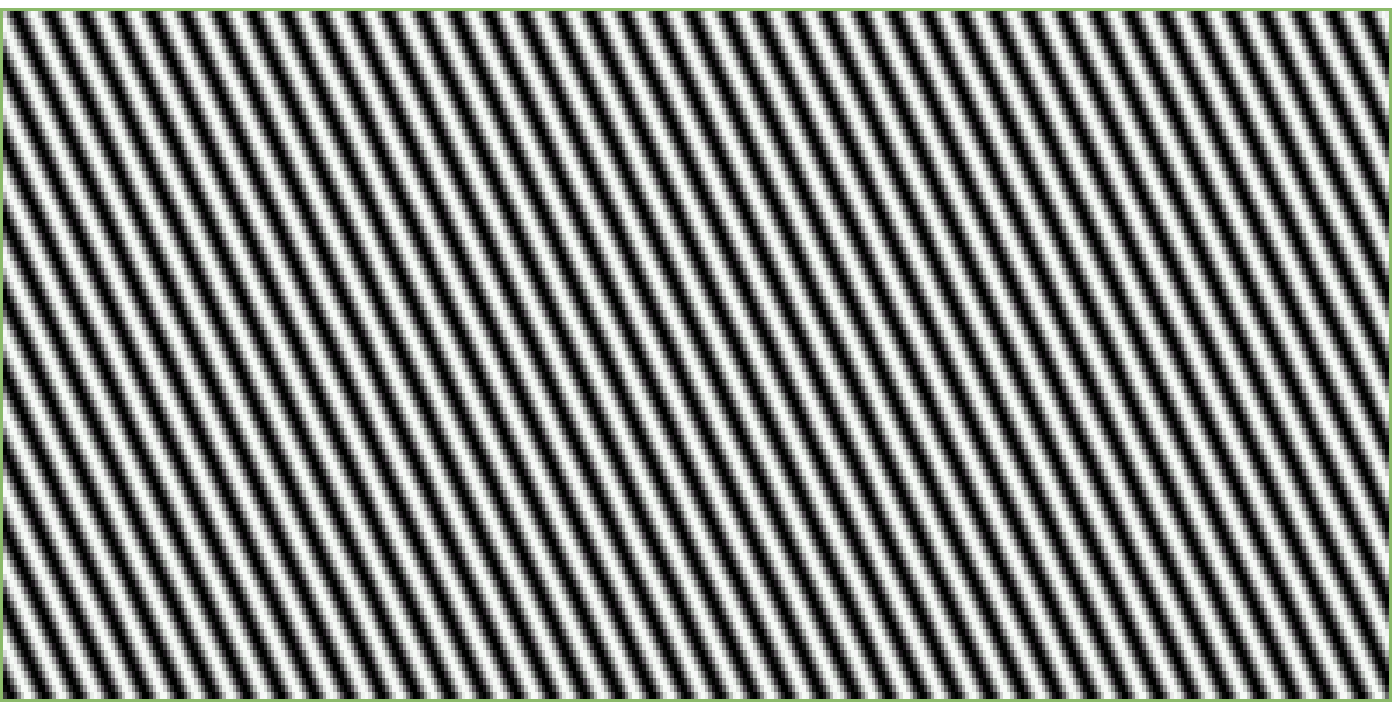}
\epsfxsize = 2.0in
\epsffile{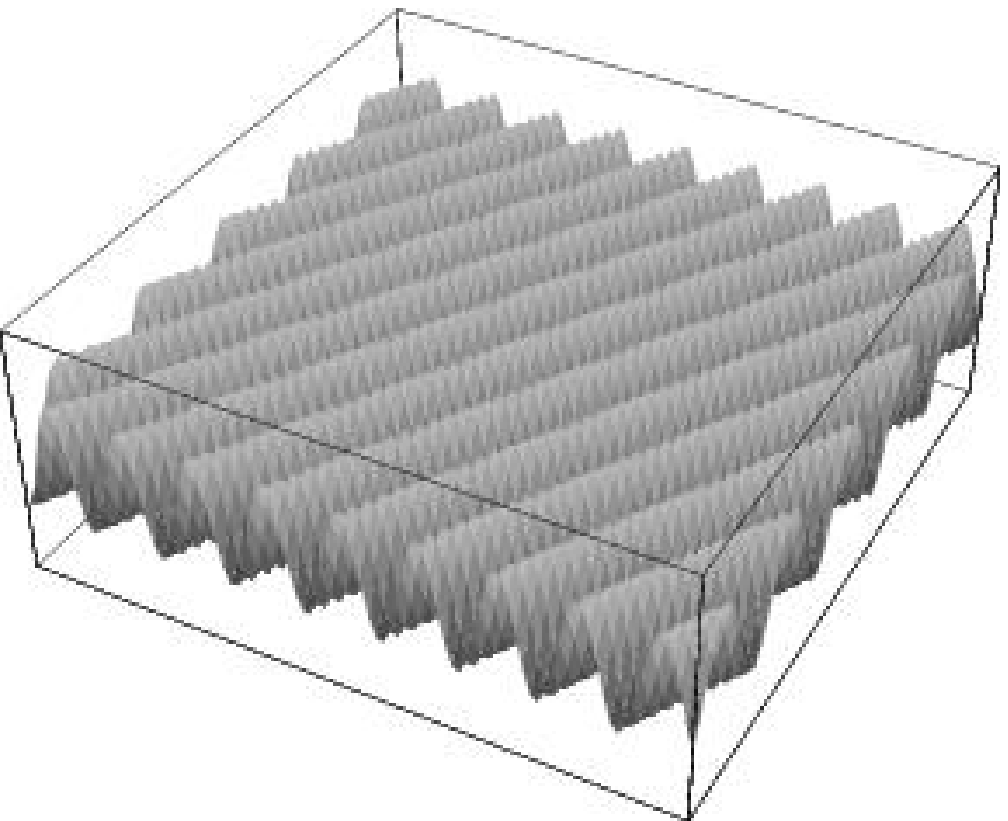}
\end{center}
\caption{Spatiotemporal behavior for the case of an unstable
membrane in region I of the pase diagram Fig.~\ref{figdiagfas2}
and the following set of parameters: $\alpha=\beta=\gamma=1$,
$\phi_0=0$, $\kappa=0.5$, $H_0=0.2$,
$\Gamma=0.14$ and $\xi=5$ (dashed curve in Fig.~\ref{figomega}).
The two upper panels correspond to the $\phi$- and
$h$-field distributions at $t=15000$,
once the structures are robust. The stripes travel from left to right
at constant velocity.
In the third panel we plot the temporal evolution (horizontal axis) of
a one-dimensional cross-section $\phi$-profile
(vertical axis) at $y=50$, starting at $t=15000$ until $t=19000$.
In the fourth panel, the corresponding $h$-profile
(horizontal axis) is plotted against time (downward below sheet axis),
starting at $t=15000$ until $t=15500$.
Darker (lighter) regions are richer in the $A$ ($B$) lipid in
the concentration snapshots,
and some exaggeration along the vertical direction
has been applied to the height plots.}
\label{figtrav}
\end{figure*}

\begin{figure}[htb]
\begin{center}
\includegraphics[width=8cm ]{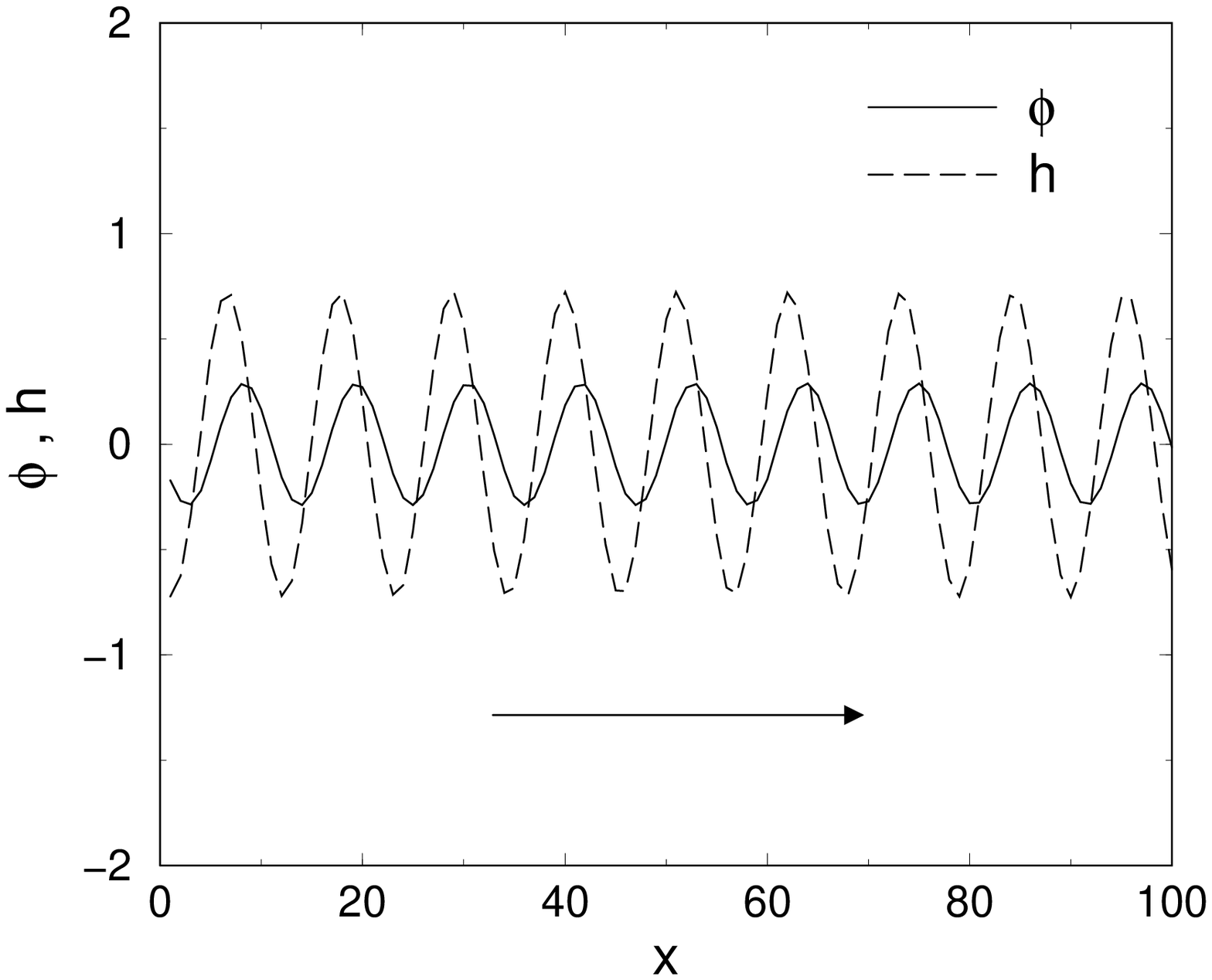}
\caption{$\phi$- and $h$-profiles for the case in Fig.~\ref{figtrav} at a given time $t=18000$.
The arrow represents the direction of propagation of the traveling pattern.}
\label{figperf}
\end{center}
\end{figure}

However, in order to accurately delimit the range of applicability
of our model, it is useful to add some comments regarding the value
of $\kappa$. The elastic properties, and thus the bending rigidity,
of lipid bilayers are strongly determined by the size, shape and
molecular elasticity of their constituents.
Membranes composed of phospholipids (which have two hydrophobic chains)
are characterized by a bending rigidity of the order of tens of
$k_BT$~\cite{kapa}.
Cell membranes containing large molar fractions of cholesterol
(rather rigid molecules)
display even higher rigidities~\cite{lenon}. In our model, the typical sizes of the
predicted dynamical patterns emerging from a wave instability for such large values
of $\kappa$ exceed the experimentally accessible sizes.
Moreover, for $\kappa \gtrsim 10$, wave-like unstable modes are observed to
be supressed at long times by the other existing Turing-like
unstable modes, so that only stationary structures are
asymptotically generated in this parameter region.
However, single- and/or short-chain lipid surfactants are known to form much
more flexible bilayers, with bending rigidities of the order of
$k_BT$ or even less~\cite{lowkapa}.
Furthermore, recent theoretical models~\cite{lowklip} show how
bilayers composed of surfactants of rather distinct shapes (as in our case)
may exhibit smaller rigidities than one-component membranes.
This, therefore, delimits the range of applicability of the model
results in this paper.

Our quantitative assigment of $\xi$ has been made with the energetic
feasibility of a number of external sources in mind.  To see this, note that
the characteristic time for a single reactive event is of
order $1/\Gamma$. During this time interval the change in the height
of the membrane due to the contribution of the reactive term is of
order $~\xi$ [see. Eq.~(\ref{hvar}); $\phi-\phi_0$ is of $O(1)$].
Moreover, the local energetic ``cost" of such a height increase due to
the reactive term is of order $2\kappa\xi^2/(\Delta x)^2$
[see Eq.~(\ref{fgen})].
In most of our simulations we have used $\xi=5$ (in simulation length
units), which, for $\kappa=0.5$, corresponds
to an energy cost of order $25$ (in units of $k_BT$).
The dephosphorylation of an ATP releases about $50$kJ/mol
or $10^{-19}$J/ATP, which at room temperature is precisely of order
$25k_{B}T$. If the energy source is light, the power required to produce
$25k_{B}T$ within a time interval of order $\Gamma^{-1}$
for $\Gamma=0.14$ as used in our simulations is
only approximately $0.3\times 10^{-6}$mW/cm$^2$, which is much lower
than the power produced by typical commercial light sources used, for
example, in photosensitive Langmuir monolayer experiments. 
While some of the energy provided by these external sources (ATP, light,
etc.) would be used for processes other than the purely elastic motion
of the membrane, including dissipation into the thermal surroundings and
the movement of other masses, this order of magnitude estimate shows
that the mechanism is feasible and provides a basis for the value of the
parameter $\xi$ chosen for our simulations.

For a highly flexible bilayer, typical phase diagrams are those of
Fig.~\ref{figdiagfas2}.  As noted earlier,
region I corresponds to a regime where at least one
of the unstable modes has nonzero imaginary part,
so both types of instablity are present. Region II is the 
unstable phase also present in Fig.~\ref{figdiagfas}.
Here, the unstable modes have no imaginary parts, so that stationary
finite-sized patterns are predicted. In region III, there are no
unstable modes (stable phase).
Here we have exhibit spatiotemporal patterns associated with the new
region I.  

The nature of the spatiotemporal patterns is determined by the value of
$\phi_0$ and by the associated relative magnitudes of the amplitudes of
waves traveling in different directions. These assertions are supported
by the amplitude equation analysis of the previous section.
Thus, for a critical quench ($\phi_0=0$), the system typically
displays some transients with domains that travel in
different directions until they organize
into a coherent train of traveling stripes. The post-transient
spatiotemporal behavior is described in the different panels
in Fig.~\ref{figtrav}.  We plot the concentration and height
fields (upper panels), and one-dimensional cross-sections
showing the temporal evolution, for both order parameters.
The traveling waves are consistent with the predictions of the amplitude
equation analysis.  
Numerical profiles at a given post-transient stage reveal the mechanism
that leads to the motion of the generated spatial structures.
In Fig.~\ref{figperf} we observe that $\phi$- and $h$-profiles are slightly
displaced, the field $\phi$ being ahead in the direction of propagation.
That there is such a phase difference between the two fields is
consistent with the appearance of the contribution with an imaginary
coefficient on the right hand side of Eq.~(\ref{bpm}).

\begin{figure*}[htb]
\begin{center}
\leavevmode
\epsfxsize = 1.7in
\epsffile{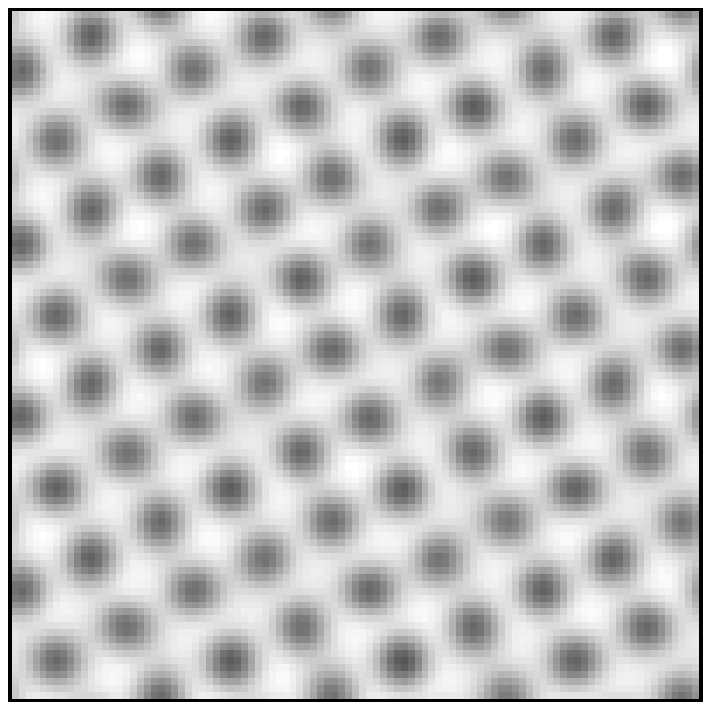}
\epsfxsize = 2.0in
\epsffile{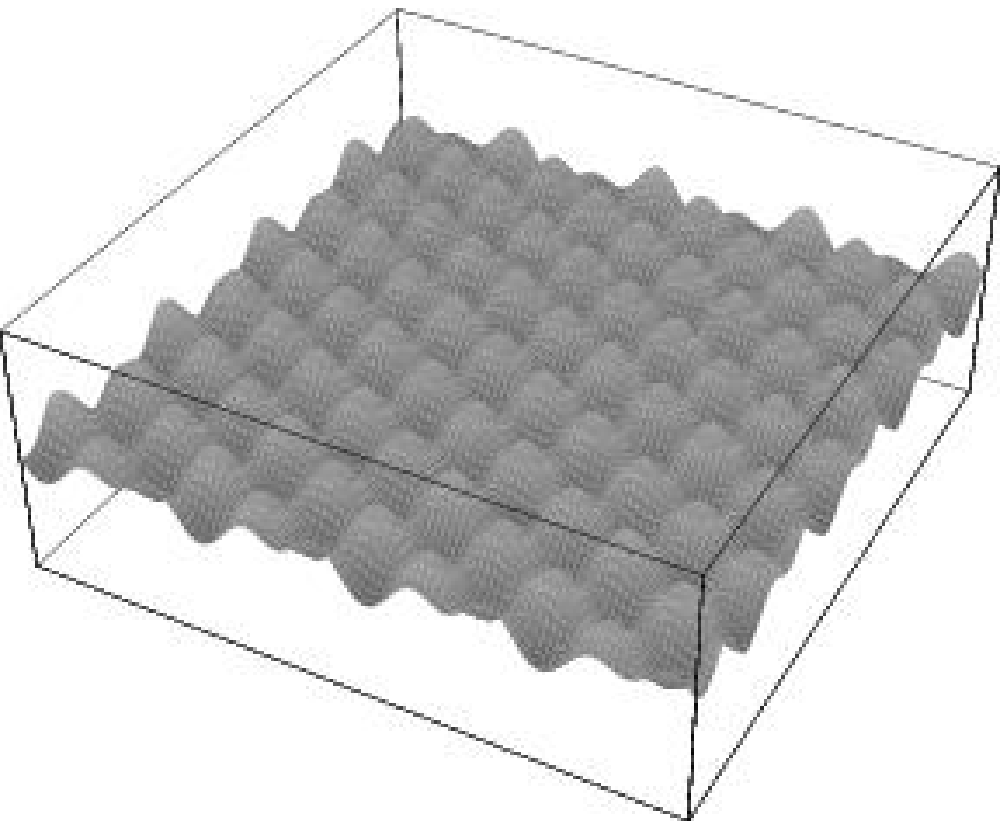} \\
\vspace{0.25in}
\epsfxsize = 2.2in
\epsffile{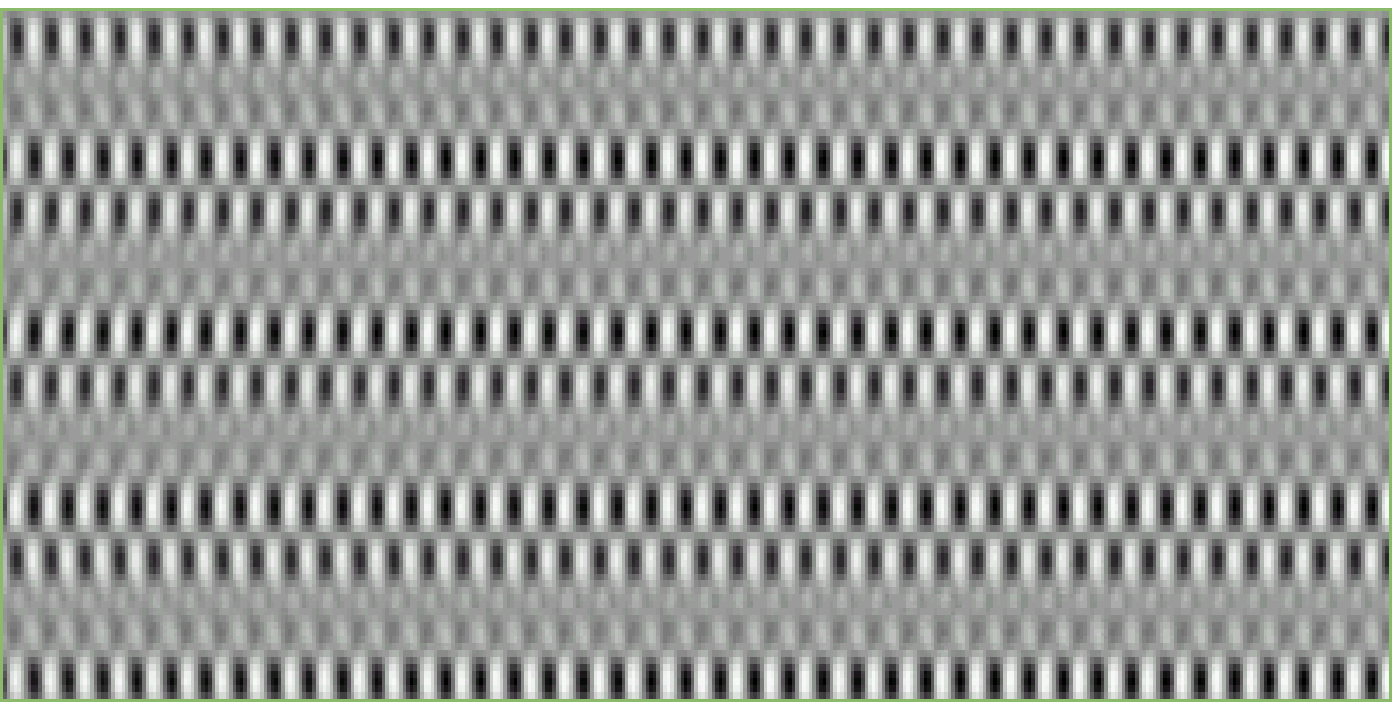}
\epsfxsize = 2.0in
\epsffile{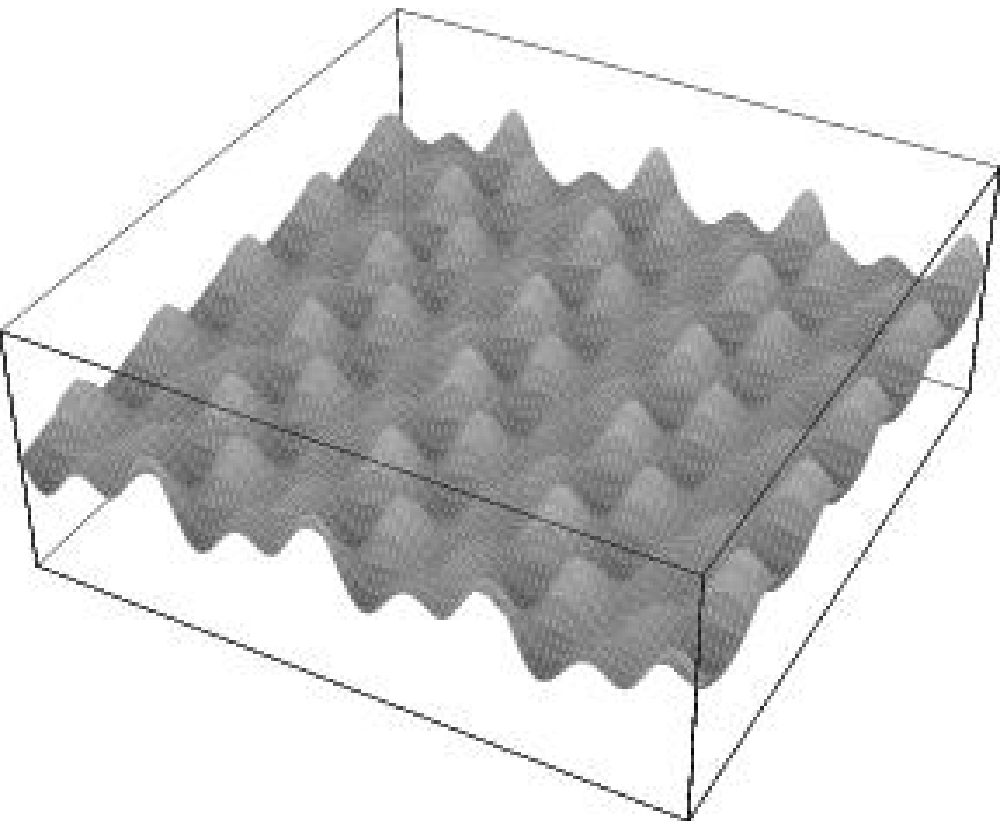}
\end{center}
\caption{Spatiotemporal behavior for the same case as in
Fig.~\ref{figtrav} except for $\phi_0=-0.14$
(still in region I). The snapshots correspond to $t=22500$ and
the temporal profile evolutions
go from this time up to $t=26500$ for $\phi$, and to $t=23000$ for $h$.
The same panel organization and specifications as in
Fig.~\ref{figtrav} apply. Bud-like curvature domains oscillate.}
\label{figoscil}
\end{figure*}

\begin{figure*}[htb]
\begin{center}
\leavevmode
\epsfxsize = 1.7in
\epsffile{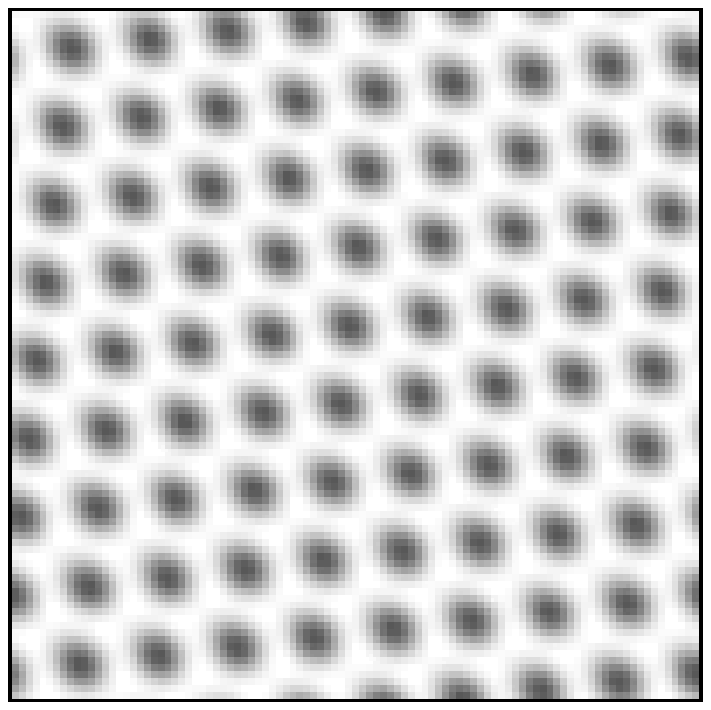}
\epsfxsize = 2.0in
\epsffile{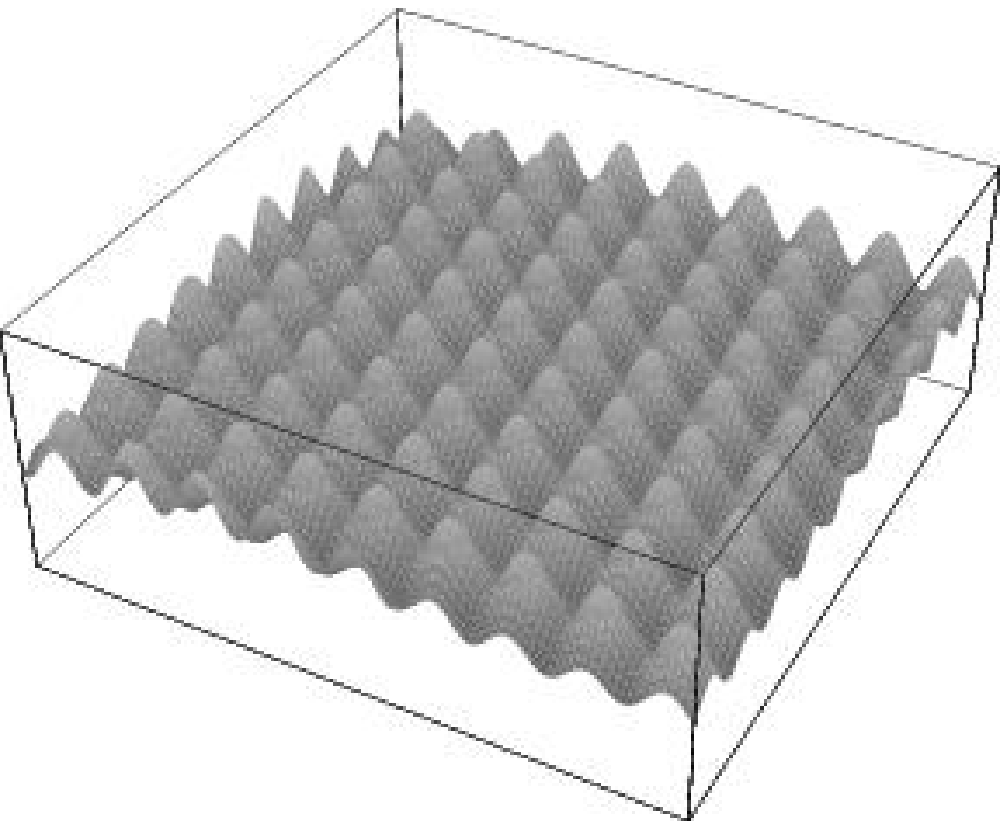} \\
\vspace{0.25in}
\epsfxsize = 2.2in
\epsffile{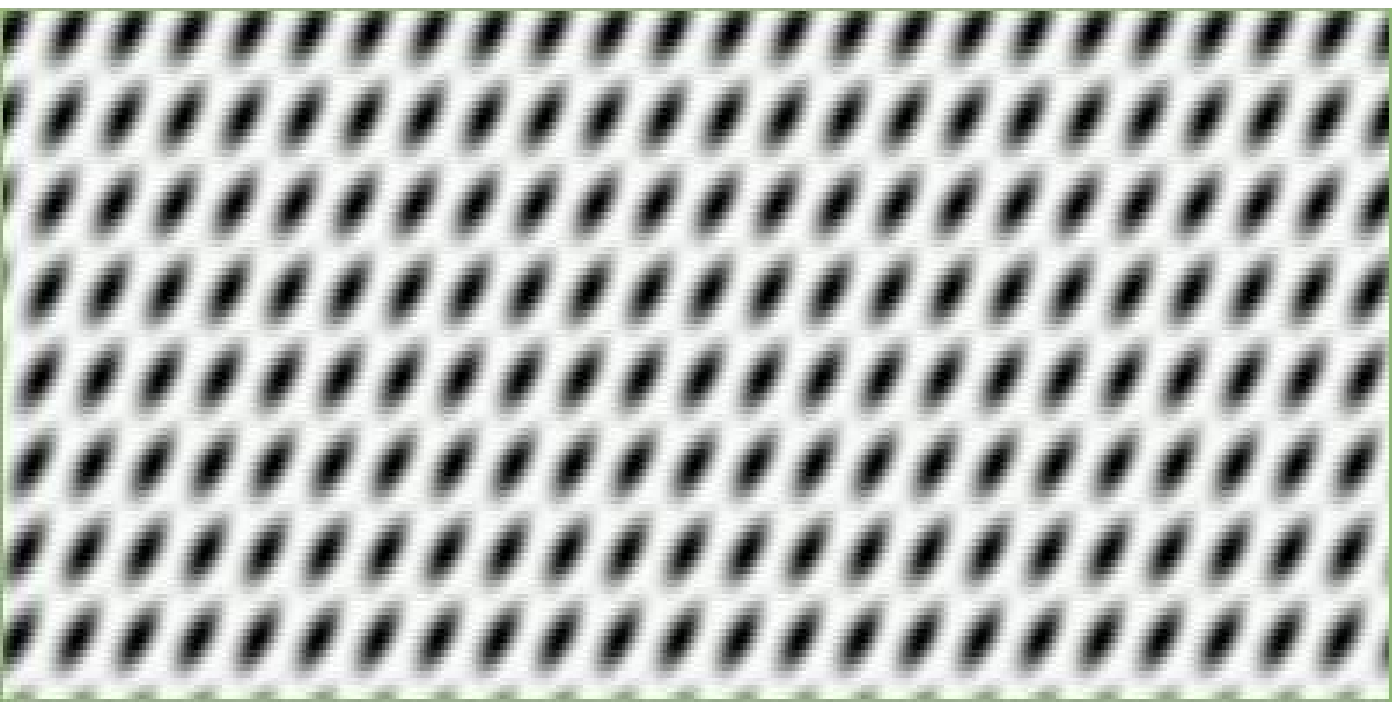}
\epsfxsize = 2.0in
\epsffile{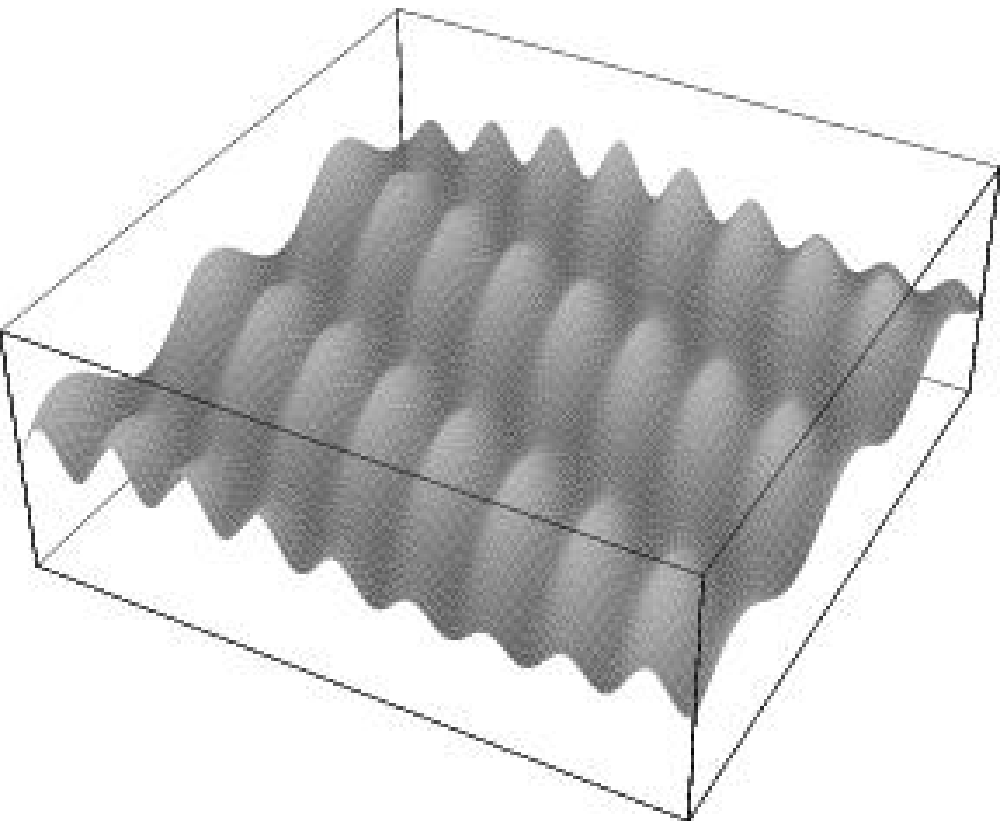}
\end{center}
\caption{Spatiotemporal behavior for the same case in
Fig.~\ref{figtrav} except for $\phi_0=-0.14$
and $\xi=3$ (still in region I). In the upper panels,
the droplet-like structures move from the upper-left corner
to the lower-right one at constant velocity, while they oscillate.
The snapshots correspond to $t=18000$ and the temporal profile evolutions
go from this time up to $t=22000$ for $\phi$, and to $t=18500$ for $h$.
The same panel organization and specifications as in
Fig.~\ref{figtrav} apply.}
\label{figmov}
\end{figure*}

Off-critical quenches ($\phi_0=-0.14$) display different
spatial and temporal behaviors, again consistent with the amplitude
equation results. As before, we first observe transients,
but now involving concentration droplets and bud-like surface deformations.
The fields settle into an oscillating pattern of buds rich in the
minority species and whose geometry depends on the value of the
parameter $\xi$ that measures the mechanical effect of the
reaction on the shape of
the membrane.  This behavior is shown in Fig.~\ref{figoscil}.
For $\xi=5$ a spatial Fourier transform of the pattern indicates
clear square symmetry.  For $\xi = 3$, the domains are rather hexagonal
and they oscillate as well as move in one direction,
as shown in Fig.~\ref{figmov}. 

\section{Conclusions}
\label{concl}

In a previous model for a flexible reactive bilayer
composed of two differently shaped lipids, we demonstrated
the occurrence of
stationary finite-sized domains of composition and curvature
as a result of the competition between thermodynamic
phase segregation and a nonequilibrium reaction~\cite{primer}.
A generalization of the model that includes the local
effect of the reaction on the membrane shape has been presented here.
We have shown how the mechanical influence of the reactive process
on the membrane may lead to the formation of spatiotemporal structures.
The linear stability analysis of the model equations shows the existence
of a wave instability for sufficiently large reaction--shape coupling.
A weakly nonlinear analysis based on the amplitude equation formalism
provides insight into the spatial and temporal symmetries of the
emerging patterns.
Correspondingly, numerical simulations display the spontaneous
generation of traveling lamelar phases, and oscillating or moving buds,
showing an important potential for the formation of spatiotemporal
patterns in deformable nonequilibrium reactive lipid membranes. 

This extension of the model should be 
considered as a formal issue in reactive deformable surfaces.
However, we believe that experimental work on giant vesicles and on
planar membranes can be designed
to qualitatively capture the spatiotemporal phenomenology obtained in
our model.
Azobenzene derivatives, which are also known to display dynamical
organization in
nonequilibrium Langmuir monolayers~\cite{our}, may be suitable
compounds to test our predictions
since their shapes can be strongly modified by means of well-known
(externally controlled) photoisomerization reactions~\cite{rau}, and
the resulting isomers normally exhibit phase separation.  

Finally, we note that it would be interesting to study the crossover
between Rouse and Zimm dynamics in our model system so as to
elucidate how the different parameters are renormalized due to
solvent hydrodynamic effects, and to clarify the resulting consequences for
the pattern formation mechanism proposed herein.
Such a study has been carried out for tethered membranes, where
different coarsening exponents are found~\cite{frey,moldovan}. 

\section*{Acknowledgments}

The authors would like to thank Dr. M. van Hecke
and a referee for fruitful comments.
Two of us (R.R and J.B.) greatfully acknowledge the
{\em Ram\'on y Cajal} program that provides
their researcher contracts. Partial support was provided by the Office
of Basic Energy Sciences at the U. S. Department of Energy
under Grant No. DE-FG02-04ER46179, and by the National Science
Foundation under Grant No. PHY-0354937.

\appendix
\section{Derivation of the amplitude equations}

For convenience of notation, we rewrite the
one dimensional version of the model equations (\ref{kineqa}) and
(\ref{kineqb}) for reactive membranes in terms of the shifted
field variable $\varphi =\phi -\phi _{0}$ and the control parameter
$\varepsilon =(\Gamma_0-\Gamma)/\Gamma_0$ that accounts for the
``distance" to the threshold of the
bifurcation to pattern formation so that a pattern develops if
$\varepsilon>0$:
\begin{subequations}
\begin{eqnarray}
\varphi_t& =&(\kappa H_0^2-\alpha) \varphi_{xx}+3\beta
(\varphi +\phi _0) \varphi_{xx}\nonumber\\
&& +6\beta (\varphi +\phi_0)\left(\varphi_x\right)^2-\gamma \varphi_{xxxx}
\nonumber\\
&& -\kappa H_0h_{xxxx}-\Gamma_0(1-\varepsilon)\varphi,
\label{eqa} \\
h_t& =&-\kappa h_{xxxx}+\kappa H_0\varphi_{xx}+\xi\Gamma_0(1-\varepsilon)
\varphi .
\label{eqb}
\end{eqnarray}
\end{subequations}
The subscripts indicate partial derivatives.  The homogeneous state
is $\varphi =0$ and arbitrary $h=\bar{h}$. With no loss of generality
we take $\bar{h}=0$.

Slightly above the bifurcation to pattern formation, the following expansion
holds (see~\cite{primer,hecke} and references therein):
\begin{subequations}
\begin{eqnarray}
\varphi & =&\sum_{n=1}^{\infty }\varepsilon^{n/2}\varphi^{(n)},
\label{4a} \\
h& =&\sum_{n=1}^{\infty }\varepsilon^{n/2}h^{(n)}, 
\label{4b}
\end{eqnarray}
\end{subequations}
and a \emph{separation of spatial scales} can be implemented between the most
unstable mode (fast growth), $q_{0}$, and the rest of the modes within
the unstable band (slower growth). In terms of the control parameter,
we let $X = \varepsilon^{1/2}x$ denote the spatial modulation scale of
the slow modes and $T=\varepsilon t$ the associated time scale. 
The separation of scales between the fastest growing mode and the slower
modes can be implemented in Eqs.~(\ref{eqa}) and (\ref{eqb}) by
expanding the spatial and temporal derivatives according to the chain
rule so that $\partial_x\rightarrow \partial_x+\varepsilon^{1/2}
\partial_X$ and $\partial_t\rightarrow \partial_t+\varepsilon\partial_T$.
Implementation of this scale separation and substitution of
Eqs.~(\ref{4a}) and (\ref{4b}) into (\ref{eqa}) (\ref{eqb}) leads to
a rather cumbersome expansion in terms of $\varepsilon$.
The lowest order contribution is of order $\varepsilon ^{1/2}$, and
balancing all terms of this order gives
\begin{subequations}
\begin{eqnarray}
&&(3\beta\phi_0^2+H_0^2\kappa -\alpha) \varphi_{xx}^{(1)}-
\gamma \varphi_{xxxx}^{(1)}-\Gamma_0\varphi^{(1)}\nonumber\\
&&~~~~~~-H_0\kappa h_{xxxx}^{(1)}-\varphi_t^{(1)}=0,
\label{7a} \\
&&\xi\Gamma_0\varphi^{(1)}+H_0\kappa \varphi_{xx}^{(1)}
-\kappa h_{xxxx}^{(1)}-h_t^{(1)} =0. 
\label{7b}
\end{eqnarray}
\end{subequations}
Equations~(\ref{7a}) and (\ref{7b}) are in fact simply the linearized
versions of Eqs.~(\ref{eqa}) and (\ref{eqb}). With the definition of
the linear operator $\mathbb{L}$ with elements 
\begin{eqnarray*}
\mathbb{L}_{11} &=& (3\beta\phi_0^2+H_0^2\kappa-\alpha)\frac{\partial^2}
{\partial x^2} -\gamma \frac{\partial^4}{\partial x^4}-\Gamma_0-
\frac{\partial}{\partial t}\\
\mathbb{L}_{12} &=&-H_0\kappa\frac{\partial^4}{\partial x^4}\\
\mathbb{L}_{21}&=&\xi\Gamma_0+H_0\kappa\frac{\partial^2}{\partial x^2}\\
\mathbb{L}_{22}&=& -\kappa \frac{\partial^4}{\partial x^4}
-\frac{\partial}{\partial t},
\end{eqnarray*}
Eqs.~(\ref{7a}) and (\ref{7b}) can trivially be written
as $\mathbb{L}\mathbf{\chi }_1=0$,
where $(\mathbf{\chi }_n)^T=\left(\varphi^{(n)},h^{(n)}\right)$.

The next order of the expansion gives us terms of $\mathcal{O}(\varepsilon)$
and yields $\mathbb{L}\mathbf{\chi}_2=\mathbf{\psi}_2\left(\left\{
\varphi ^{(1)};h^{(1)}\right\}\right)$, where
$\mathbf{\psi }_2^T=\left(\mathbf{\psi }_2^{(a)}\mathbf{,\psi}_2^{(b)}\right)$,
and
\begin{eqnarray*}
\mathbf{\psi }_{2}^{(a)}&=&-6\beta\phi_0\left[ \left(
\varphi_x^{(1)}\right)^2+\varphi^{(1)}\varphi_{xx}^{(1)}\right)\\
&&+2\left(\alpha -3\beta \phi_0^2-H_0^2\kappa\right)
\varphi_{xX}^{(1)}+4\gamma \varphi_{xxxX}^{(1)}\\
&&+4H_0\kappa h_{xxxX}^{(1)}, \\
\mathbf{\psi }_2^{(b)}&=&2\kappa \left(-H_0\varphi_{xX}^{(1)}
+2h_{xxxX}^{(1)}\right).
\end{eqnarray*}
At order $\varepsilon^{3/2}$ we get $\mathbb{L}\mathbf{\chi }_3
=\mathbf{\psi }_3\left(\left\{\varphi^{(1)},\varphi^{(2)};
h^{(1)},h^{(2)}\right\}\right)$,
where the components of $\mathbf{\psi }_3^T=\left(\mathbf{\psi }_3^{(a)}
\mathbf{,\psi }_3^{(b)}\right)$ are given by
\begin{eqnarray*}
\mathbf{\psi}_3^{(a)}&=&\varphi_T^{(1)}-\Gamma_0\varphi^{(1)}\\
&&-6\beta\varphi_{x}^{(1)}\left(\varphi^{(1)}\varphi_x^{(1)}
+2\phi_0\left(\varphi_X^{(1)}+\varphi_x^{(2)}\right)\right) \\
&& +\left(\alpha-H_0^2\kappa\right)\left(\varphi_{XX}^{(1)}
+2\varphi_{xX}^{(2)}\right)\\
&&-3\beta \left\{ \phi_0^2\left(\varphi_{XX}^{(1)}
+2\varphi_{xX}^{(2)}\right)\right.\\
&&\left. +\varphi_{xx}^{(1)}\left[\left(\varphi^{(
1)}\right)^2+2\phi_0\varphi^{(2)}\right] \right. \\
&&\left. +2\phi_0\varphi^{(1)}\left(2\varphi_{xX}^{(1)}
+\varphi_{xx}^{(2)}\right) \right\}  \\
&& +2\gamma \left(3\varphi_{xxXX}^{(1)}+2\varphi_{xxxX}^{(2)}\right)\\
&&+2H_0\kappa\left(3h_{xxXX}^{(1)}+2h_{xxxX}^{(2)}\right) , \\
\mathbf{\psi }_{3}^{(b)}&=&h_T^{(1)}+\xi \Gamma_0\varphi^{(1)}
+\kappa \left[2\left(3h_{xxXX}^{(1)}+2h_{xxxX}^{(2)}\right) \right.\\
&&\left.-H_{0}\left( \varphi_{XX}^{(1)}+2\varphi_{xX}^{(2)}\right)
\right]. 
\end{eqnarray*}

We could continue up to any order with the expansion. However, at order
$\varepsilon^{3/2}$ we are able to extract a closed evolution equation for
the amplitudes of the pattern, as shown below, and we therefore stop
at this order. Note that
beyond order $\varepsilon ^{1/2}$, which corresponds to the linear problem,
in all cases we obtain a non-homogeneous equation, such that at order
$\varepsilon^{n/2}$
\begin{equation}
\mathbb{L}\mathbf{\chi }_n=\mathbf{\psi }_n\left( \left\{ \varphi
^{(1)},\ldots ,\varphi^{(n-1)};h^{(1)},\ldots, h^{(n-1)}\right\} \right). 
\label{ecuaciones}
\end{equation}
At order $\varepsilon^{1/2}$ the problem is homogeneous and,
with appropriate boundary conditions, we can write the solution as
\begin{subequations}
\begin{eqnarray}
\varphi^{(1)}&=&A_{1+}(X,T)e^{i(q_0x+\omega_0 t)}\nonumber\\
&&+A_{1-}(X,T)e^{i(q_0x-\omega_0 t)}+c.c.,
\label{12a} \\
h^{(1)}&=&B_{1+}(X,T)e^{i(q_0x+\omega_0 t)}\nonumber\\
&&+B_{1-}(X,T)e^{i(q_0x-\omega_0 t)}+c.c.,
\label{12b}
\end{eqnarray}
\end{subequations}
where $c.c.$ stands for the complex conjugate and
$\omega_0=\left\vert {\rm Im}(\omega_{q_0})\right\vert$.
However, the amplitudes $A_{1\pm }$ and $B_{1\pm }$ are
undetermined at this point. As mentioned above, the
subsequent orders are no longer homogeneous and therefore the existence of
an analytical solution can not be ensured. However, we can enforce
solvability by evoking the \emph{Fredholm alternative
theorem}~\cite{newell}.
In our case the application of the theorem simply states, as a recipe,
that for Eqs.(\ref{ecuaciones}) to have a solution, the
functions $\psi_n$ can not contain the fundamental mode
$\exp \left[\pm\left(iq_0 x\pm\omega_0 t\right)\right]$. In other words,
the particular solutions of the
non-homogeneous problem are orthogonal to the solutions of the homogeneous
problem. Thus, by substituting the solutions (\ref{12a}) and (\ref{12b})
into the next order
of the hierarchy and imposing the solvability condition we obtain the
following particular solution,
\begin{eqnarray*}
\varphi^{(2)}&=&c_{1\varphi}\left[A_{1+}(X,T)\right]^2
e^{2i(q_0 x+\omega_0 t)}  \\
&&+c_{2\varphi}\left[A_{1-}(X,T)\right]^2
e^{2i(q_0 x-\omega_0 t)} \\
&&+c_{3\varphi}A_{1+}(X,T) A_{1-}(X,T)e^{2iq_0 x} +c.c., \\
h^{(2)}& =&c_{1h}\left[\left(A_{1+}(X,T)\right)^2
e^{2i(q_0 x+\omega_0 t)}  \right. \\
&& \left. +\left(A_{1-}(X,T)\right)^2
e^{2i(q_0 x-\omega_0 t)}\right] \\
&&+c_{2h}A_{1+}(X,T)A_{1-}(X,T) e^{2iq_0 x} +c.c. ,
\end{eqnarray*}
where
\begin{eqnarray*}
c_{1\varphi}&=& \frac{c_{1\varphi}^{num}}{c_{1\varphi}^{den}}\\
c_{1\varphi}^{num}&=& \left[12\beta \phi_0q_0^2\left(8\kappa
q_0^4+i\omega_0\right) \right]\\
c_{1\varphi}^{den}&=&
\left[8H_0\kappa q_0^4(4H_0\kappa q_0^2-\Gamma_0\xi)\right.  \\
&&-(8\kappa q_0^4+i\omega_0)\left(\Gamma_0+4q_0^2(3\beta
\phi_0^2+H_0^2\kappa \right. \\
&& \left. \left. +4\gamma q_0^2-\alpha) +2i\omega_0\right) \right], \\
c_{2\varphi } &=&-\frac{\left[12\beta \phi_0q_0^2(8\kappa q_0^4-
i\omega_0)\right]}{c}, \\
c_{3\varphi } &=&-\frac{24\beta \phi_0q_0^2}{\Gamma_0(1+H_0\xi)
+4q_0^2(3\beta\phi_0^2+4\gamma q_0^2-\alpha) }, \\
c_{1h} &=&\frac{\left[6\beta\phi_0q_0^2\left(-\xi \Gamma_0+4H_0
\kappa q_0^2\right) \right]}{c^{\ast} }, \\
c_{2h} &=&-\frac{3\beta \phi _0q_0^2(\xi \Gamma_0-4H_0\kappa q_0^2)}
{2\kappa q_0^4\left[ \Gamma_0(1+H_0\xi) +4q_0^2(3\beta \phi_0^2+4\gamma
q_0^2-\alpha) \right] },\\
c &=&8\kappa q_0^4\left[\Gamma_0(1+H_0\xi)
+4q_0^2(3\beta\phi_0^2+4\gamma q_0^2-\alpha) \right]  \\
&&-2\omega _0^2 -i\omega_0\left[\Gamma_0 \right. \\
&&\left. +4q_0^2\left( 3\beta
\phi_0^2+H_0^2\kappa -\alpha +4q_0^2(\gamma +\kappa)\right) \right],
\end{eqnarray*}
and $c^\ast$ is the complex conjugate of $c$.
The values of $A_{1\pm}$ and $B_{1\pm}$ can not be determined at
this order either.  However, at order $\varepsilon ^{3/2}$ the
Fredholm theorem provides closed
equations for the conditions that must be satisfied by
these amplitudes.
These conditions are the amplitude equations for our pattern
forming system, and are given in terms of the original variables
$x$ and $t$ in Eqs.~(\ref{ampa}) and (\ref{ampb}).  In those equations
the ``1" in the subscript has been dropped for economy of notation.


\end{document}